\documentclass[12pt,aps]{revtex4}
\usepackage{epsfig,amsmath,amssymb}

\newcommand{\bea}{\begin{eqnarray}}
\newcommand{\eea}{\end{eqnarray}}

\newcommand{\dbar}{d\hskip -0.4em ^-}

\begin{document}

\title{Electric dipole moment of the neutron from \\
 a  flavor changing Higgs-boson}

\vspace{0.5cm}

\author{Jan O. Eeg}

\email{j.o.eeg@fys.uio.no}
\affiliation{Department of Physics, University of Oslo,
P.O.Box 1048 Blindern, N-0316 Oslo, Norway}

\vspace{0.5 cm}

\begin{abstract}

I consider neutron electric dipole moment contributions induced by  
flavor changing Standard Model Higgs boson couplings to quarks. Such
 couplings might stem
 from non-renormalizable $SU(2)_L \times U(1)_Y$ invariant 
Lagrange terms of dimension six,  containing a
 product of three Higgs doublets.
 Previously  one loop diagrams with
such couplings were considered in order to constrain
 {\it quadratric} expressions of  Higgs flavor
 changing couplings to quarks.
In the present paper the analysis is extended to the two  loop level, where
there are diagrams for electric dipole moments of quarks 
 with a flavor
 changing Higgs coupling to {\it first order only}.
 The divergent loops, due to non-renormalisabillity, are 
parametrized in terms of an ultraviolet  cut-off $\Lambda$.
I also consider
QCD corrections, including the mixing with the color electric dipole moment,
while the contribution from the Weinberg operator is found to be negligible.
 The effect of QCD corrections  is to suppress the bare result.

Using the current
experimental bound on the neutron electric dipole moment, then
for cut offs from one to seven TeV, I find a 
 constraint of order $10^{-3}$ for the imaginary part of the
{\it product} of the Higgs flavor changing coupling for
 $(d \rightarrow b)$-transition   {\it and} the 
CKM element $V_{td}$.
 Assuming that the previous bound of the {\it absolute value} of the 
 Higgs flavor changing coupling for $(d \rightarrow b)$-transition
 obtained from $B_d - \bar{B_d}$-mixing is saturated,
 the experimental bound on the neutron electric dipole moment would be 
reached  for the {\it bare} result, {\it if} the cut off were extended up to
about ca 20 TeV. However, QCD corrections suppress this result by a
 factor of order ten, and keep the nEDM below the experimental bound.

\end{abstract}

\maketitle

\newpage


\section{Introduction}
 
An electric dipole moment (EDM) for elementary particles is a CP-violating
 quantity and it
gives important information on the matter anti-matter asymmetry 
in the universe. 
EDMs of elementary fermions within the Standard Model (SM) are  induced
 through the Cabibbo-Kobayashi-Maskawa (CKM) CP-violating
phase. EDMs are studied also within  many models
 Beyond the SM (BSM). For
reviews on SM and BSM EDMs,
 see\cite{Pospelov:2005pr,Fukuyama:2012np,Dekens:2014jka,Jung:2013hka,Yamanaka:2017mef}.
 Experimentally  only 
 bounds on
electron, muon, proton and neutron EDMs are  determined \cite{Olive:2016xmw}.
 Explicitly, for the EDM of
 the neutron (nEDM $ \equiv \, d_n$) discussed in this paper, the present
 experimental bound is
\cite{Baker:2006ts} 
\begin{equation}
d_n^{exp}/e   \le 2.9 \times 10 ^{-26} \,  \mbox{cm} \; .
\label{dn-bound}
\end{equation}

 Within the SM, the nEDM is calculated to be several orders of magnitudes below 
the experimental
bound. Calculations of the  nEDM will in general put bounds 
on hypothetical models BSM,
 and any measured nEDM significantly bigger that the SM estimate 
($10 ^{-32}$ to  $10 ^{-31} \, e$~cm)
 would signal New Physics.

The SM contributions to the nEDM are well known and thoroughly 
explained in \cite{Pospelov:2005pr}. 
At a low energy scale one can construct 
an effective Lagrangian
\begin{equation}
{\cal L}_\mathrm{eff} \; = \;{\cal L}_4 + {\cal L}_5 + {\cal L}_6 + ... \, ,
\label{ToteffLag}
\end{equation}
with all possible CP-odd operators of appropriate dimension. 
The QCD-odd term gives
the dimension 4 operator \cite{Pospelov:2005pr,Yamanaka:2017mef}.
 The dimension 5 term contains  electric dipole
moment operators  as well as  color electric operators of quarks. 
The color electric operator and the the CP-odd three-gluon Weinberg
operator (of dimension
 six) will in general mix under QCD renormakization 
\cite{Pospelov:2005pr,Fukuyama:2012np,Dekens:2014jka}.
The electric dipole moment of a single fermion in (2) has the form
\begin{equation}
{\cal L}_\mathrm{5,em} \; = \; \frac{i}{2} \, d_f \,
\bar{\psi_f} \sigma_{\mu \nu} \, F^{\mu \nu} \, \, \gamma_5 \psi_f
\; \; ,
\label{effLag}
\end{equation}
where $d_f$ is the electric dipolement of the fermion,
$\psi_f$ is the fermion (quark) field, $F^{\mu \nu}$ is the electromagnetic 
field tensor,  and
$\sigma_{\mu \nu} = i [\gamma_\mu, \, \gamma_\nu]/2$ is the dipole operator
 in Dirac space. The color electric dipole operator is given by the same 
expression with $d_f$ replaced by the color electric dipole moment $d^c_f$ 
and $F_{\mu \nu}$ replaced by $G^a_{\mu \nu} \, t^a$, where $G^a_{\mu \nu}$
 is the color octet gluon tensor, and $t^a$ are the $SU(3)_c$ color matrices.
 The electric dipole operator in (\ref{effLag}) for quarks
 appears within the SM
 from three loop 
diagrams with double Glashow-Iliopoulos-Maiani (GIM)-
cancellations and in addition a gluon exchange. These are of
 order $\alpha_s G_F^2$, and
are proportional to quark masses
and an imaginary Cabibbo-Kobayashi-Maskawa(CKM) factor. They
  were found to be very small, of order
$10^{-34} e $ cm \cite{Shabalin:1980tf,Czarnecki:1997bu}.
Still within the SM, many contributions to the nEDM due to 
 interplay of quarks in
 the neutron, were studied
\cite{Pospelov:2005pr,Maiezza:2014ala,Nanopoulos:1979my,Morel:1979ep,Gavela:1981sk,Khriplovich:1981ca,McKellar:1987tf,Eeg:1982qm,Eeg:1983mt,Hamzaoui:1985ri,Mannel:2012qk}.
These mechanisms, gave results of order $10^{-33}$ to $10^{-31} e$ cm.

The nEDM due to EDMs of light $u$- and $d$-, and even $s$-quarks
 may be given by the formula
\begin{equation}
d_n \, = \, \rho_u \, d_u \; +  \; \rho_d \, d_d \;  + \; \rho_s \, d_s 
\; ,
\label{valencedn}
\end{equation}
similar to a corresponding   formula for the magnetic moment.
In the strict valence approximation, 
\begin{equation}
 \rho_u \,= - \frac{1}{3} \;  ,  \; \, \rho_d \, = \; \frac{4}{3} \; ,
  \; \, \rho_s \, = 0 \;  ,
\label{valencecoeff}
\end{equation}
while  lattice calculations 
\cite{Bhattacharya:2015esa,Bhattacharya:2015wna} give 
\begin{equation}
 \rho_u \,= - 0.22  \pm 0.03 \; ,  \; \, \rho_d \, = \; 0.74 \pm 0.07 \;
  , \; \,  \rho_s \, = 0.008 \pm 0.010\; . 
\label{valencelatt}
\end{equation}
Note that there is a contribution to the nEDM from the EDM of
 the $s$-quark, with a small coefficient.

Many models  BSM  suggest possible
  new particles and/or new interaction Lagrange terms 
inducing  EDMs  \cite{Pospelov:2005pr,Fukuyama:2012np,Dekens:2014jka,Jung:2013hka,Yamanaka:2017mef,Maiezza:2014ala,Buchmuller:1982ye,Manohar:2006ga,Altmannshofer:2010ad,Buras:2010zm,Degrassi:2010ne,Arnold:2012sd,Brod:2013cka,He:2014uya,Fajer:2014ara,Fuyuto:2015ida,Bian:2016zba}. In
 the case of New Physics (NP) presence, flavor physics might be testable through
CP-violating asymmetries in mesonic decays
\cite{Altmannshofer:2010ad,Buras:2010zm,Altmannshofer:2012ur}. 
The properties and couplings
of the physical Higgs boson ($H$) are still not completely known. 
 Some authors
 \cite{Goudelis:2011un,Blankenburg:2012ex,Harnik:2012pb,Greljo:2014dka,Gorbahn:2014sha,Dorsner:2015mja} have suggested that the
physical Higgs boson might have flavor changing  couplings to 
fermions which might also
be CP-violating. In these papers bounds on quadratic expressions of 
such couplings 
 were  obtained  from various processes, 
say, like $K-\bar{K}, \,D-\bar{D} \,$, and $ B-\bar{B}$ - mixings, and 
also from leptonic flavor 
changing decays like $\mu \rightarrow e \, \gamma$
 and $\tau \rightarrow \mu \, \gamma$. In the latter case 
 two loop diagrams of Barr-Zee type \cite{Barr:1990vd} were also considered
\cite{Goudelis:2011un,Blankenburg:2012ex,Harnik:2012pb,Chang:1993kw}.
(See also \cite{Leigh:1990kf}).
Flavor changing couplings of this type will occur
if the SM Higgs  have non-renormalized interactions appearing when 
higher mass states are integrated out.
For instance, 
flavor changing Higgs (FCH) couplings might stem from
  $SU(2)_L \times U(1)_Y$-invariant but 
non-renormalizable Lagrangian terms of dimension six.

The purpose of the present paper is to  extend the
 analysis of \cite{Blankenburg:2012ex,Harnik:2012pb}
 to two loop diagrams.
In the one loop case one needed two FCH couplings to generate the EDM.
In the two loop case it is however possible to find diagrams with the FCH 
coupling {\it to first order only}, while the rest of the couplings are
 ordinary SM couplings.

Some of these two loop diagrams considered here give contributions 
suppressed by the 
small mass ratio $m_d/M_W$ 
for the ordinary SM Higgs coupling to fermions. ($M_W$ denotes the mass of the
 $W$-boson and $m_q$ is the mass of the quark $q$).
However, if the Higgs is coupled to a top ($t$) quark
 one might obtain
relevant non-suppressed contributions. Motivated by the result of
 the previous 
work \cite{Fajer:2014ara},  I consider such diagrams.
There are additional reasons to extend the analysis in  
\cite{Blankenburg:2012ex,Harnik:2012pb} for nEDM to two loop 
level. 
Namely, in general, it is known that some two loop diagrams might give
 bigger amplitudes 
than one loop diagrams
because of  helicity flip(s) in the 
latter
\cite{Blankenburg:2012ex,Harnik:2012pb,Chang:1993kw,Bjorken:1977vt}. 
In the present case,  two loop amplitudes will be proportional 
to a large $ttH$ coupling
or a large $WWH$ coupling within the SM, in contrast to the small SM Higgs
 couplings to
light fermions. This might compensate for the two loop suppression of 
the diagrams. I have also adressed the issue of perturbative QCD corrections,
 which turn out to suppress the bare result. 

In the next section (II) I will present the framework for the FCH
 couplings. In the  sections III and IV  two loop
 calculations for the FCH couplings will be presented. The QCD corrections are presented in section V.
  In section  VI
the results will be discussed, and  the  conclusion  given in section VII.
An Appendix is given in section VIII. 

\section{Flavor Changing Physical Higgs?} 
Within the framework in  \cite{Goudelis:2011un,Blankenburg:2012ex,Harnik:2012pb,Greljo:2014dka,Gorbahn:2014sha,Dorsner:2015mja} (see also ref. \cite{Giudice:2008uua})
 the effective interaction Lagrangian for the FC transition
 $f_1 \rightarrow f_2$ due to Higgs exchange can in general be written
\begin{equation}
  {\cal L}_\mathrm{eff} \, = \, 
Y_R(f_1 \rightarrow f_2) \cdot  \overline{(f_2)_L} \, H \, (f_{1})_R \, 
 +  \; h.c. \; ,
\label{FCNCf}
\end{equation}
where $f_{1,2}$ are fermion fields, $H$ the physical Higgs field and 
 $Y_{R}(f_1 \rightarrow f_2)$'s are 
 coupling constants,  thought to be complex numbers. 
 Then,
  from the hermitean conjugation
 part, there will be a left-handed   $f_2 \rightarrow f_1$
coupling
\begin{equation}
 Y_R(f_1 \rightarrow f_2)^{*} \, = \, Y_L(f_2 \rightarrow f_1) \; \;
\label{complex-rel}
\end{equation}

 Flavor changing Higgs couplings of the type presentes in 
eq. (\ref{FCNCf})  may occur if there are non-renormalizable  
Higgs type Yukawa-like interactions due to dimension six operators, as 
shown explicitly in \cite{Harnik:2012pb,Dorsner:2015mja} :
\begin{equation} 
  {\cal L}^{(D)}  \, =  \, - \, \lambda_{ij} \, \overline{Q_i} \Phi  
\, D_j \, - \, \frac{\tilde{\lambda}_{ij}}{\Lambda_{NP}^2} 
 \, \overline{Q_i} \Phi  D_j \, (\phi)^\dagger \Phi \, + \; h.c. \; ,
\label{FCNClad}
\end{equation}
where the  generation indices $i$ and $j$ 
are understood to be summed over the values 1,2,3.
Further,  $\Phi$ is the SM Higgs field, $Q_i$ is the left-handed $SU(2)_L$ 
quark doublets, and the  $D_j$'s are the right-handed $SU(2)_L$ singlet 
$d$-type quarks in a general basis. 
 Moreover, 
$\Lambda_{NP}$ is the scale where New Physics is assumed to appear.
There is a similar term as in (\ref{FCNClad}) for 
  right-handed type $u$-quarks, $U_j$.

\begin{center}
\begin{figure}[htbp]
\scalebox{0.6}{\includegraphics{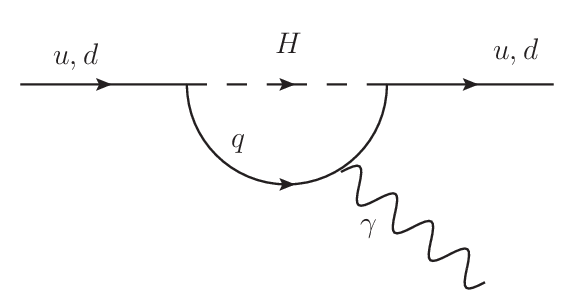}}
\caption{One loop diagrams for EDMs of $u$- and $d$-quarks 
with FC Higgs couplings. Here $q=s,b$ for an EDM of the $d$-quark and 
$q=c,t$ for an EDM of a $u$-quark.}
\label{one-loop}
\end{figure}
 \end{center}

Using the assumptions based on  (\ref{FCNCf}), one obtains one loop
 diagrams for 
EDMs of $u$- and $d$-quarks  \cite{Blankenburg:2012ex,Harnik:2012pb}.
The one loop diagram in Fig. \ref{one-loop} 
- with FCH coupling at both vertices, puts bounds on {\it quadratic}
 expressions of the $Y$'s for definte choices of flavor. Note that this
 diagram gives a finite contribution to quark EDMs.

\section{Diagrams with one FC coupling -and a $t \bar{t} H$-coupling}
\begin{center} 
\begin{figure}[htbp]
\scalebox{0.55}{\includegraphics{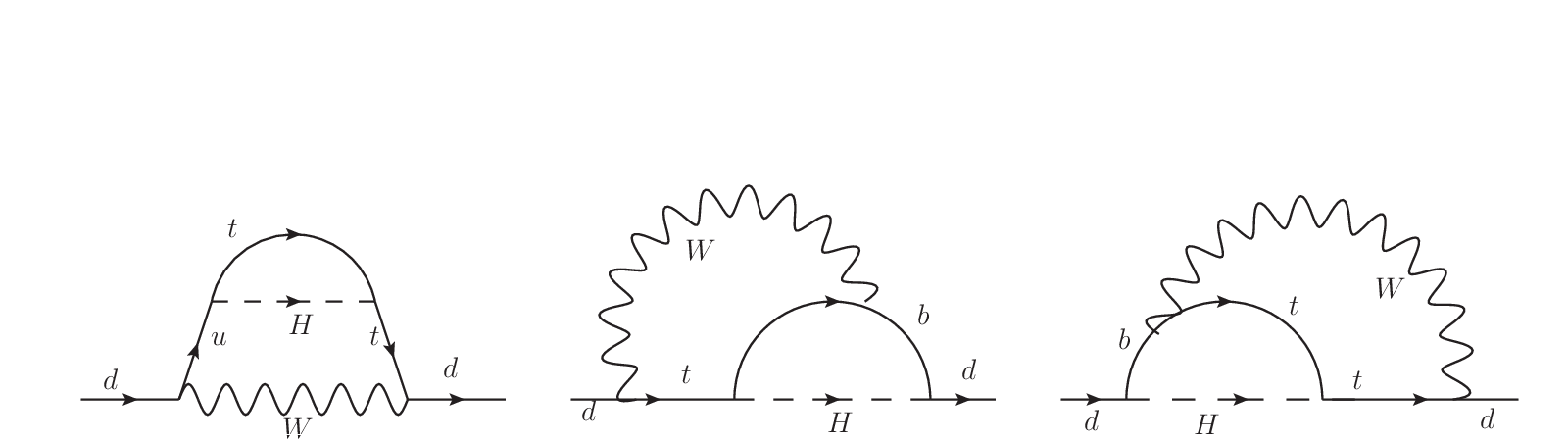}}
\caption{Three diagrams   with FC Higgs coupling for EDMs of
a $d$-quark. Soft photon emission from one of the charged particles 
is assumed to   be added. The left diagram will give  contributions 
suppressed by $m_{u,d}/M_W$.
Taking the crossed diagrams in the center or to the right, we will get
 contributions which are not suppressed by light quark masses.
 The  diagram to the right  is the  complex conjugate
 of the  diagram in the middle. 
}
\label{FCHNew2loop}
\end{figure}
 \end{center}

\begin{center}
\begin{figure}[htbp]
\scalebox{0.70}{\includegraphics{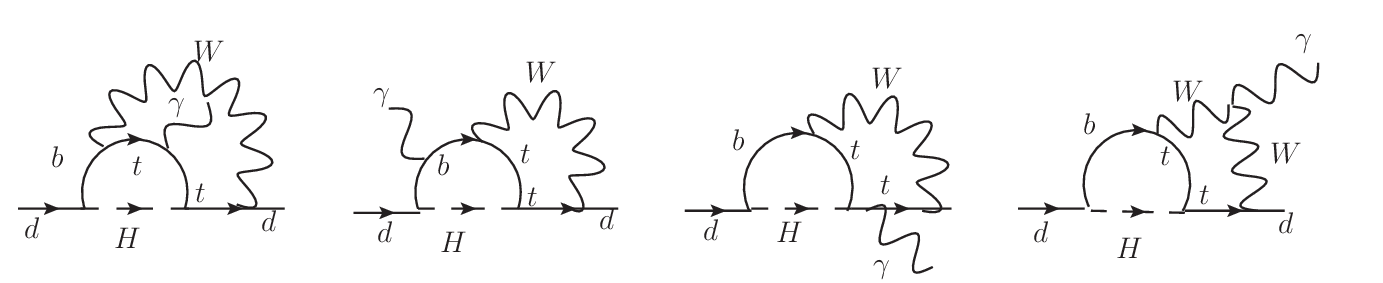}}
\caption{Four diagrams for an EDM of a $d$ quark obtained by adding a
 soft photon to the diagram to the right of Fig. 2.
There are also corresponding diagrams 
where the $W$-boson is replaced by an unphysical Higgs-boson within
 Feynman gauge.
 }
\label{FCHNew2loopg}
\end{figure}
 \end{center}

In Fig. \ref{FCHNew2loop} are  shown some  two loop 
diagrams   for 
the EDM of a $d$-quark generated by exchange of one physical Higgs ($H$)
 boson and one $W$-boson, with a sizeable Higgs coupling
 $\sim g_W \, m_t/M_W$ to a top quark and 
where  only {\it one} of the Higgs couplings are flavor changing
(a soft  photon is assumed to be added).
The non-crossed
 version to the left in Fig. \ref{FCHNew2loop} does not give  
non-suppressed contributions. Taking 
crossed Higgs and $W$-bosons are equivalent to the topologies in the middle 
and right of Fig. \ref{FCHNew2loop}.
 Adding a soft
  photon to the diagram in the middle
and to the right,  we get four diagrams for both cases.
In Fig.  \ref{FCHNew2loopg} the four diagrams obtained by adding a soft 
photon emission to the diagram to the right in Fig. \ref{FCHNew2loop} are shown.

I have found that the results for the loop contributions
 in Fig. \ref{FCHNew2loopg}
 have the form: 
\begin{eqnarray}
{\cal M}(f \rightarrow f \gamma)_{(a)} \; =
 \;  A \, \left(\bar{f} \, \sigma \cdot F \, P_R \, f \right)
  \, \, ,
\label{eq:Two-loop}
\end{eqnarray}
and that the diagrams with interchanged order of $H$ and $W$ loops,
 as in the middle of 
Fig. \ref{FCHNew2loop}
have the form:
\begin{eqnarray}
{\cal M}(f \rightarrow f \gamma)_{(c)} \; = 
 \;   A^* \, \left(\bar{f} \, \sigma \cdot F \, P_L \, f \right)
  \, ,
\label{eq:Two-loopC}
\end{eqnarray}
where $P_L = (1-\gamma_5)/2$ and $P_R = (1+\gamma_5)/2$ are  projectors
 in Dirac space.
Thus the  electric dipole moment is found to be:
\begin{eqnarray}
 (d_f)_{2-loop} = 2 \, Im(A) \; . 
\label{dfA}
\end{eqnarray}
There is also a contribution to the magnetic moment (i.e the gyromagnetic
 quantity $(g-2)$) given by  $2 \, Re(A)$.

The contributions from the four diagrams ($i =$ 1-4) in
Fig. \ref{FCHNew2loopg} and its complex conjugates
can  then, by using (\ref{dfA}) be written 
\begin{eqnarray}
 (\frac{d_d}{e})_i \, =  \hat{e}_i \, F_2 \, S_i \, Im[Y_R(d \rightarrow b) \,
 V_{td}^* \, V_{tb}] \; ,
\label{eq:Amp1}
\end{eqnarray}
where the $\hat{e}_i$'s are the electric charges (in units $e$= the
 proton charge) of the photon-emitting particles, i.e.   
$\hat{e}_{1,3}= \hat{e}_t = + 2/3$,  $\hat{e}_{2}= \hat{e}_b = - 1/3$, and 
$\hat{e}_{4}= \hat{e}_W = +1 $.
 Here I have used the relations (\ref{complex-rel}) and (\ref{dfA}).
Note that a left-handed coupling $Y_L(d \rightarrow b) \, P_L$ would not 
contribute in (\ref{eq:Amp1}) due to wrong chirality.
 The $V$'s are CKM matrix 
elements in the standard notation.
The constant $F_2$ sets the overall scale of the EDMs obtained from the
 two loop diagrams:
\begin{equation}
F_2 \, = \,  \frac{g_W^3}{M_W \, \sqrt{2}} \left(\frac{1}{16 \pi^2}\right)^2
\, = \, \frac{2 M_W^2}{v^3} \left(\frac{1}{16 \pi^2}\right)^2 \,
 \simeq \, 6.94 \times 10^{-22} \, \mbox{cm} \; ,
\label{F-quantity}
\end{equation}
where I have used the conversion rule $1/(200MeV)=10^{-13}$cm.
The quantities $S_i$  in (\ref{eq:Amp1}) are  dimensionless functions
 of the masses of the particles entering the two loop diagrams.
 Some details from the loop calculations are 
given in the Appendix.

Using Feynman gauge for the $W$-boson, one has also to add diagrams with
the unphysical Higgs field 
 $\phi_\pm$ ({\it i.e.} the longitudinal component of the 
$W$-boson) given by the 
 the Lagrangian 
\begin{equation}
  {\cal L}_\mathrm{\phi t d} \, = \, 
\; - \frac{g_W}{M_W} \, V_{td}^* \bar{d} \, \phi_- (m_d P_L \, - \,
 m_t P_R) t \, + \, h.c. \; .
\label{phi-ferm}
\end{equation}

 For  finite loop diagrams, a typical example is given in 
 (\ref{loopint})-(\ref{loopFunc}), while other finite integrals  are given
 with same type formulae with permuted masses,
The loop functions
$S_1$ and $S_2$ are finite, dimensionless, and depend on the mass ratios 
\begin{equation}
u_t = (m_t/M_W)^2 =4.64 \;\; ,  \quad 
u_H = (M_H/M_W)^2=2.44   \; .
\label{mass-ratios}
\end{equation}
The masses of the $W$-boson, the $t$-quark and the physical Higgs-boson $H$
are of the same order of magnitude. Therefore, because of  lack of a clear mass
 hierarchy, it makes no sense to consider leading logarithmic approximations,
 in contrast to \cite{Fajer:2014ara}.
Numerically, I find
\begin{equation}
S_1 = \, - \,  2.55  \; \; \; ; \; \; S_2 =  2.50 \; \, .
\label{S12Num}
\end{equation}

If the soft photon is emitted from the top quark after exchange of
 the Higgs boson, as in the third diagram from left
  in Fig. \ref{FCHNew2loopg}, or  
from the $W$-boson in the fourth diagram, the left sub-loop containing
 the Higgs boson 
is logarithmically divergent, which is not unexpected because the interaction in
(\ref{FCNClad})  is non-renormalizable.
Each of the divergent integrals $\sim ln(\Lambda^2)$ are followed  by  finite
 logarithmic terms more cumbersome than for finite loop integrals,
 and such integrals are given
 by expressions like in
(\ref{N-funct}), also with masses permuted for different diagrams.

The total contribution from the third digram in  Fig. \ref{FCHNew2loopg},
including the contribution
 from the 
unphysical Higgs, is
\begin{equation}
S_3 \, = \, u_t \, p_1(u_t) \, C_\Lambda + 1.81 \; .
\label{S3Tot}
\end{equation} 
Here  the UV divergence is
 parametrized through the quantity
\begin{equation} 
C_\Lambda \;\equiv  \; ln(\frac{\Lambda^2}{M_W^2}) \, + \, \frac{1}{2} \; \, ,
\label{P-CLambdiv}
\end{equation}
where $\Lambda$ is the UV cut-off. Numerically,  $C_\Lambda $ is $\sim 5.5$
to 9.4 for $\Lambda \sim $ 1 to 7 TeV. 
Furthermore, $ u_t \, p_1(u_t)$ is the result of the second subloop. Here
\begin{equation}
p_1(u) \, \equiv \, \frac{u}{(u-1)} \left(1 - \frac{ln(u)}{u-1} \right) \; \;  ;
\quad p_1(u_t)= 0.737 \; ,
\label{p1-func}
\end{equation}
where $u_t$ is given in (\ref{mass-ratios}).

The fourth diagram  in Fig. \ref{FCHNew2loopg} with the soft photon  emitted 
from the $W$-boson again contains
 a divergent part, and 
the total contribution to the fourth diagram  is
\begin{equation}
S_4 \, = \, - \frac{3}{4} \, p_2(u_t) \, C_\Lambda + 2.98 \; ,
\label{S4Tot}
\end{equation}
where
\begin{equation}
p_2(u) \, \equiv \,
 \frac{u}{(u-1)} \left( \frac{u \cdot ln(u)}{u-1} \, - 1 \right) 
\; \; ; \quad p_2(u_t)\, = \, 1.219 \; .
\label{p2-func}
\end{equation}

Summing all contributions from diagrams in Fig. \ref{FCHNew2loopg}, I find 
\begin{eqnarray}
 (\frac{d_d}{e})_{Fig.3} \, =  \, (1.65 \, + \, 1.37 \, C_\Lambda) \cdot  F_2 \; 
 \cdot Im[Y_R(d \rightarrow b) \, V_{td}^* \, V_{tb}] \; .
\label{eq:AmpIII}
\end{eqnarray}

 There are in addition contributions from the same diagrams in 
 Fig. \ref{FCHNew2loopg}, but with other quarks in the loop.
If the $b$-quark is replaced by an $s$-quark, the CKM factors are two orders
of magnitude smaller, and in addition $Y_R(d \rightarrow s)$ has a
 stricter bound
from $K -\bar{K}$-mixing. If the $t$-quark is replaced by 
the $u$- or $c$-quark, the contributions  are suppressed
 by $(m_u/m_t)^2$ and $(m_c/m_t)^2$, respectively.

There are also similar diagrams for EDM of an $u$-quark, {\it i.e.} like
 in Fig. \ref{FCHNew2loopg} with the $t$- and the $b$-quarks interchanged. This
 amplitude has the same structure as in (\ref{eq:Amp1}), and is proportional to
the combination 
$Im[Y_R(u \rightarrow t) \cdot V_{tb}^* \, V_{ub}]$. But the $u$-quark EDM
 contributions will be neglected. First,
the ordinary SM coupling of the Higgs will be proportional to 
$m_b/M_W$ instead of  $m_t/M_W$ for the $d$-quark case.
Then it turns out that the prefactors $S_i$ for $u$-quark EDM contributions
 are  suppressed by a factor of order $(m_b/m_t)^2 \sim 10^{-3}$
 compared to the analogous $d$-quark contributions. 
Second,  even if the ratio between
 $Y_R(u \rightarrow t)$ and $Y_R(d \rightarrow b)$ would be of order 
$m_t/m_b$, the $u$-quark EDM contribution to the nEDM in  (\ref{valencedn})
 would still be suppressed by
$|(\rho_u \cdot m_b)/(\rho_d \cdot m_t)| \,  \sim 10^{-2}$ compared to the
$d$-quark EDM contribution to the nEDM.

\section{Diagrams with one FC coupling -and a $WWH$-coupling}
\begin{center}
\begin{figure}[htbp]
\scalebox{0.55}{\includegraphics{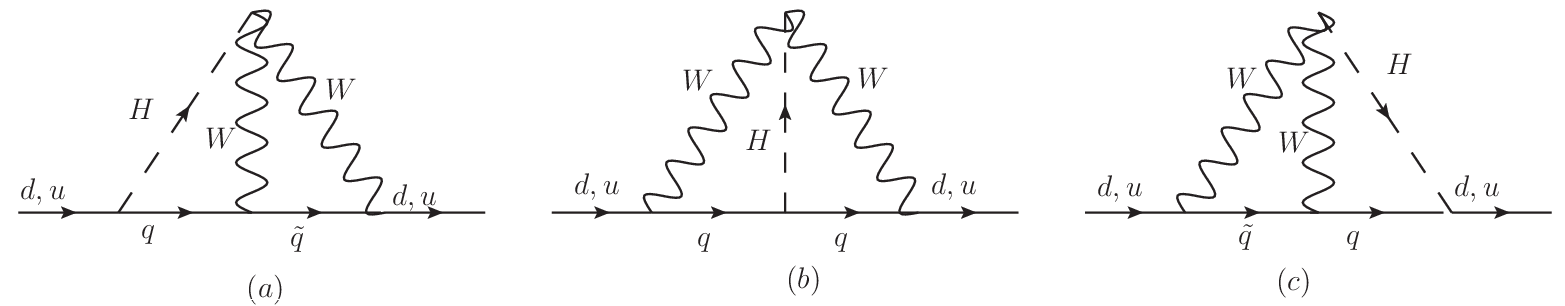}}
\caption{Three diagrams with one $WWH$-Higgs coupling and one
 FC Higgs coupling for EDMs of
 a $u$- or $d$-quark. Soft photon emission from one of the charged particles 
is assumed to   be added.  The (b) diagrams are
 zero in the limit of zero external momentum of the light quarks due
 to momentum integration, or are suppressed by small light quark masses.
 The (c) diagrams are  complex conjugates
 of the (a) diagrams . Here $q=s,b$ and $\tilde{q}=u,c,t$ for 
EDM of a $d$-quark, and  $q=c,t$ and $\tilde{q}=d,s,b$ for EDM of a
 $u$-quark. }
\label{FCH2loop}
\end{figure}
 \end{center}

\begin{center}
\begin{figure}[htbp]
\scalebox{0.6}{\includegraphics{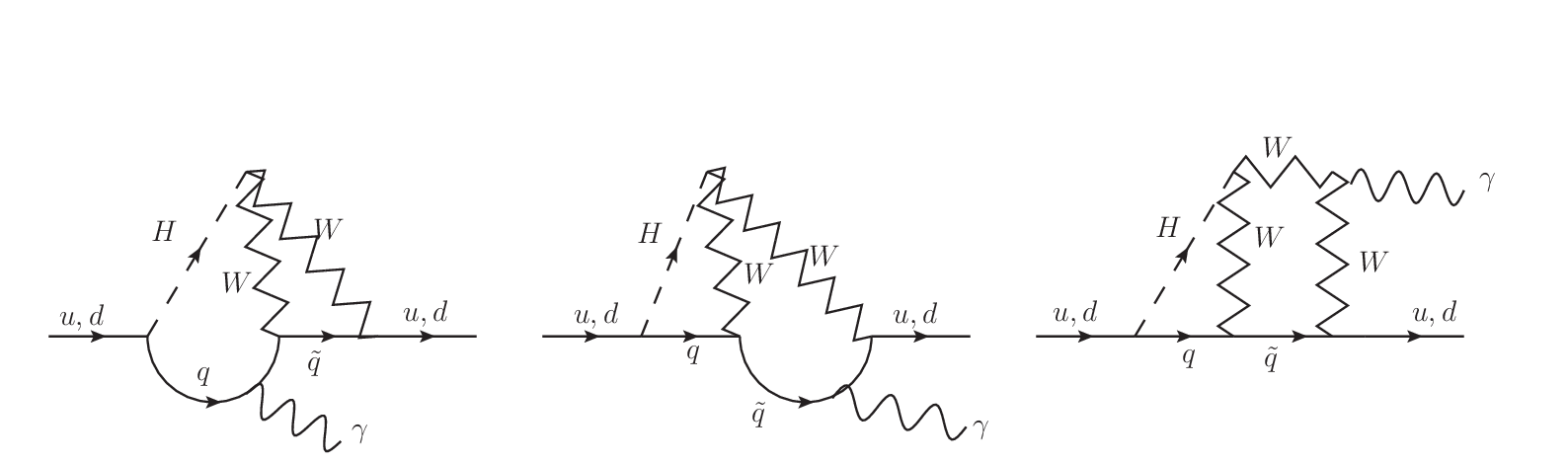}}
\caption{Emission of a soft photon from (a)-type diagrams of 
Fig. 4. 
There is also a diagram with emission from the $W$ in center of the diagram, 
and in addition  graphs with the  W replaced
 by an unphysical Higgs within Feynman gauge.}
\label{2LoopWWHgam}
\end{figure}
 \end{center}

We will now consider another class of two loop diagrams 
 generated by FC Higgs-boson  couplings.
 These diagrams shown in  Fig. \ref{FCH2loop}  
have a big  $WWH$-coupling $\sim g_W \, M_W$ and {\it only  one}
 FC Higgs coupling to a fermion.
These two loop diagrams are divided in three types: the (a)-diagrams
 with Higgs exchange to the left, the (b)-diagrams with Higgs exchange in
 the middle, and the (c)-diagrams with Higgs exchange to the right.
In the limit of small external light quark momenta, which we work,
   the (b)-diagrams are 
 zero due to (odd) momentum integration, or they are suppressed by small 
external quark masses.
 The (c)-diagrams are  complex conjugates
 of the (a)-diagrams.
 Soft photon emission from one of the charged particles  should of course
     be added in Fig. \ref{FCH2loop}, as seen in Fig. \ref{2LoopWWHgam}
 for the (a)-diagrams. The (a) diagrams give contributions like 
in (\ref{eq:Two-loop}), and the (c) diagrams like in  (\ref{eq:Two-loopC}).
The relevant piece of the SM  Lagrangian for a Higgs coupling to two $W$-bosons
  is given by
\begin{equation}
  {\cal L}_\mathrm{WWH} \, = \,g_W \, M_W \, \sqrt{2} \, H \, 
W^{(-) \, \mu} \, W_\mu^{(+)} \; .
\label{WWH}
\end{equation}
Using Feynman gauge for the $W$-boson, we must also consider Lagrangian terms
for a physical Higgs coupling to a $W$-boson and the
 unphysical Higgs boson $\phi_\pm$. 
 In addition to the term
 for quarks coupling to $\phi_\pm $ in (\ref{phi-ferm}),
 there is  the relevant $H W \phi_\pm$-
coupling
  obtained from the Lagrangian 
\begin{equation}
  {\cal L}_\mathrm{HW\phi} \, = \, \frac{g_W}{\sqrt{2}}
 \left\{ \, H \, (i \partial^\mu \phi_-) \, - 
\, (i \partial^\mu \, H) \phi_- \right\} \, W_\mu^{(+)} \, + \,  h.c.
\label{phi}
\end{equation}
Because of derivative couplings,
 the vertices involving the  unphysical Higgs $\phi_\pm$ will 
depend on the loop momenta, which might give  divergent (sub-)loops.
There are also  $W \gamma \phi_\pm$-couplings, 
but they do not contribute for soft photon emission.

In the preceeding section (III), for all the shown diagrams in 
Fig.\ref{FCHNew2loopg}, the physical Higgs coupled to the top quark
with strength $\sim g_W m_t/M_W$. Also the chiral structure of the diagrams
 is such that these diagrams are  
proportional to $m_t^2$, and even $m_t^4$ in $S_3$. In the present section
the diagrams have a flavor blind $W W H$ coupling, and have another
 chiral structure,  
 and one gets diagrams $\sim \, m_t^2$ only for the case when 
the $W$-boson is replaced by an unphysical 
Higgs $\phi_\pm$. Therefore I have apriori considered all quark flavors in the
 loops, although it is expected that the GIM-mechanism will cancel 
the leading terms with  light quark flavors, except for the difference
 between the $t$-quark and the $c$-quarks contribution.

Contributions to the $d$-quark EDM from soft photon  emission from
 the quark $q= s,b$ in   
the diagram 5a,( {\it i.e.}
 to the left in Fig. \ref{2LoopWWHgam}) can in the general case be written :
\begin{eqnarray}
 (\frac{d_d}{e})_{5a} \, = \,- \, 2 \, \hat{e}_s \, F_2 \, \left\{ 
\mbox{Im} \left[Y_R(d \rightarrow s) \, \lambda_u\right] \cdot \Delta f_d(s,u-c) 
+ \mbox{Im} \left[Y_R(d \rightarrow s) \lambda_t \right] \cdot \Delta f_d(s,t-c)
\right\}
\nonumber \\ 
- \, 2 \, \hat{e}_b \,  F_2 \, \left\{
 \mbox{Im}\left[ Y_R(d \rightarrow b) \, \xi_u \right] \cdot \Delta f_d(b,u-c)
+ \mbox{Im}\left[ Y_R(d \rightarrow b) \, \xi_t \right] \cdot \Delta f_d(b,t-c)
 \, \right\} \, ,
\label{dFCHfCompl}
\end{eqnarray}
where $F_2$ is given in (\ref{F-quantity}) and 
where $\hat{e}_s = \hat{e}_b = -1/3$ are charges for the photon-emitting
quarks, and  the $\lambda$'s and the $\xi$'s are CKM factors:
\begin{equation}
\lambda_{\tilde{q}} = V_{\tilde{q} d}^*  V_{\tilde{q} s} \; , \qquad 
\xi_{\tilde{q}} = V_{\tilde{q} d}^*  V_{\tilde{q} b} \; , \qquad 
 \tilde{q} = u,c,t \; \, .  
\label{CKMProd}
\end{equation}

The $\Delta  f$'s in (\ref{dFCHf}) are differences, due to GIM-cancellation,
between loop functions  $f(q,\tilde{q})$, for given flavors $q =s,b$ and
 $\tilde{q} =u,c,t$. These are functions of  quark, $W$-boson
 and Higgs masses.
Above, I have used the shortages
\begin{equation}
\Delta f_d(b,t-c) \; \equiv \;  f(b,t) \, - \, f(b,c) \quad \; ;
\; \; \Delta f_d(s,c-u) \; \equiv \;  f(s,c) \, - \, f(s,u) \; \, , 
\label{GIMf}
\end{equation}
and so on in a self-explanatory way. 

The quantities $f(q,\tilde{q})$  are finite,
and  are of order $10^{-1}$ to $1$. The GIM-cancellations
for the difference between the $u$- and $c$-quark contributions are very 
efficient, such that the differences $\Delta f_d(s,u-c)$ and 
 $\Delta f_d(b,u-c)$ are of order $10^{-3}$ to $10^{-7}$,
and can be safely neglected. 
Contributions with the $t$-quark in the loops are significantly
 different from contributions involving the lighter quarks.
Thus, the determination
of {\it both} the  $t$-quark {\it and} the $c$-quark contributions
 will be important.
 In this case  
the GIM cancellation is not efficient. 

Contributions to the $d$-quark EDM from soft photon  emission from
 the quark $q= s,b$ in   
the diagram 5a,the dominating contribution can be written :
\begin{eqnarray}
 (\frac{d_d}{e})_{5a} \, = \, 
- \, 2 \, \hat{e}_b \,  F_2 \, \left\{
 \mbox{Im}\left[ Y_R(d \rightarrow b) \, \xi_t \right] \cdot \Delta f_d(b,t-c)
 \, \right\} \, ,
\label{dFCHf}
\end{eqnarray}
where $F_2$ is given in (\ref{F-quantity}) and 
where $\hat{e}_b = -1/3$ is the charge for the photon-emitting
$b$-quark, and  $\xi_{\tilde{q}} = V_{td}^*  V_{t b}$.
There are also other contributions which are small and can be 
neglected.

The quantity $\Delta f_d(b,t-c)$ from loop calculations is the
 difference between the $b \rightarrow t$ and the $b \rightarrow c$ 
contributions, and is given by
\begin{equation}
 \Delta f_d(b,t-c) \simeq -0.34 \; .
\label{Dfd-value1}
\end{equation}

The diagram with  soft photon emission from the quark $\tilde{q} = u,c,t$,
 is shown the center of Fig. \ref{2LoopWWHgam} ({\it i.e} Fig. 5b).
Adding contributions where
 the $W$ is replaced by an unphysical Higgs $\phi_\pm$, one  obtains 
  divergent contributions for these loop functions.

 Because the $WH\phi_\pm$-vertex is momemtum dependent, 
  the left subloop is divergent,
 reflecting again that  
the theory based on the Lagrangian in eq. (\ref{FCNCf}) alone is not 
renormalizable.
The numerically relevant term
 from diagram  5b is given by
\begin{eqnarray}
 (\frac{d_d}{e})_{5b} \, \simeq  \, + 2 \, \hat{e}_{\tilde{q}} \, 
 F_2 \cdot 
\mbox{Im} \left[ \xi_t \, Y_R(d \rightarrow b) \right] \cdot
 \Delta h_d(b,t-c) \; \, ,
\label{dFCHh-dom}
\end{eqnarray}
where $\Delta h_d(b,t-c)$ is defined similar to 
the $\Delta f$'s in eq.(\ref{GIMf}).
In this case there is a divergent term  when $W$ is replaced by 
the unphysical Higgs $\phi_\pm$, and the total result from diagram 5b is
\begin{equation}
\Delta h_d(b,t-c) \, = \, p_2(u_t) \, C_\Lambda - 1.26 \; ,
\label{h-term}
\end{equation}
where $C_\Lambda$ is given in (\ref{P-CLambdiv}) and $p_2(u)$ in (\ref{p2-func}).

An example for diagrams with a soft photon emitted from the $W$-boson is
  shown at the right of  Fig. \ref{2LoopWWHgam} (Fig. 5c).
Also in this case there are divergent
 diagrams, because the left sub-loop might be divergent for the replacement
$W \rightarrow \phi_{\pm}$. 
After GIM-cancellation the dominant term is
\begin{eqnarray}
 (\frac{d_d}{e})_{5c} \, \simeq \, 3 \, \hat{e}_W \,   F \, \cdot 
\mbox{Im} \left[ Y_R(d \rightarrow b) \, \xi_t \right] \cdot \Delta k_d(b,t-c)
 \,  \, ,
\label{dFCHW-dom}
\end{eqnarray}
where 
  one finds
\begin{equation}
\Delta k_d(b,t-c) \, = \, \frac{1}{2} p_2(u_t) \, C_\Lambda - 2.48
\label{k-term}
\end{equation}

Neglecting small contributions (all except those proportional to 
 $V_{td}^* \, V_{tb} \, \equiv \xi_t$), and 
summing all contributions 
  from diagrams in Fig. \ref{2LoopWWHgam}
one finds 
\begin{eqnarray}
 (\frac{d_d}{e})_{Fig.5} \, =  \, (3.46 \, C_\Lambda - 9.35) \cdot F_2 \cdot
 Im[Y_R(d \rightarrow b) \, V_{td}^* \, V_{tb}]
\label{eq:AmpIV}
\end{eqnarray}
The EDM of the $u$-quark is neglected due to small loop functions 
(-after GIM-cancellation), small CKM-factors. Moreover, the comments about the
 $Y_R$'s at the end of the previous sections are also relevant here. 

\section{Perturbative QCD corrections}

Summing all two loop contributions from section III and IV, I obtain the
total bare dominanting contribution for an EDM of the $d$-quark:
\begin{eqnarray}
 (\frac{d_d}{e})_{Tot}^{bare} \, =  \, \equiv C_E(\mu_\Lambda)(4.83 \, = \,
 C_\Lambda - 7.70) \, F_2 \,  \,
 Im[Y_R(d \rightarrow b) \, V_{td}^* \, V_{tb}] \; .
\label{eq:AmpTot}
\end{eqnarray}
But perturbative QCD effects must also been taken into account.
The color electric term can be easily found from the same expressions 
for photon emission from quarks (corresponding to  all quark charges put to
$+g_s$).
The total color electric term is then found to be
\begin{eqnarray}
 (\frac{d_d}{g_s})_{Tot}^{bare} \, =  \, C_C(\mu_\Lambda) \, = \, 
(1.96\, C_\Lambda - 2.55) \, F_2 \,  \,
 Im[Y_R(d \rightarrow b) \, V_{td}^* \, V_{tb}] \; .
\label{eq:AmpTotQCD}
\end{eqnarray}

\begin{center}
\begin{figure}[htbp]
\scalebox{0.55}{\includegraphics{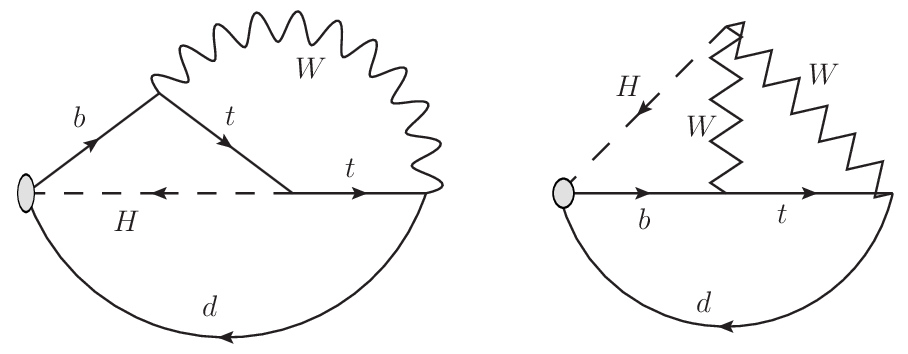}}
\caption{The contributions to the Weinberg operator for the flavor
 changing Higgs interaction $Y_R(d~\rightarrow~b)$, represented by the 
grey blobs to the left in both diagrams. Three gluon lines have to 
be attached to quark lines. 
  } 
\label{WeinbLoop}
\end{figure}
 \end{center}

There are also  contributions from the Weinberg operator for the 
FCH couplings. Contributions
 to the Weinberg operator
proportional to  $Im[Y_R(d \rightarrow b) \, V_{td}^* \, V_{tb}]$,
 are shown in Fig.~\ref{WeinbLoop}.
These are however very small
 due to ``wrong'' chiralities,  are suppressed by $m_d/M_W$, and will
 therefore be neglected.
The color electric term mixes
into the EDM term in (\ref{effLag}) due to renormalization effects 
in perturbative QCD. The relevant mixing matrix under QCD 
renormalization at one loop level  is given in \cite{Degrassi:2005zd}.
This result
 is also used in  \cite{Gorbahn:2014sha}.
The result for  the coefficient $C_E$ of
 the EDM-operator describing the running from a high scale $\mu_{high}$  down
 to a smaller scale $\mu_{low}$ is
\begin{equation}
C_E (\mu_{low}) \, = \, \eta^{\kappa_E} C_E(\mu_{high}) \, +  
\, \frac{\gamma_{CE}}{\gamma_E \, - \gamma_C} \,
 (\eta^{\kappa_E} \, - \, \eta^{\kappa_C}) \,  C_C(\mu_{high}) \; .
\end{equation}
There is also a term due to the   Weinberg operator
which is omitted here because of the negligible contribution mentioned above.
In this one loop formula,
 $\gamma_E$ and  $\gamma_C$ are  the anomalous dimensions
 of the EDM-  and the color electric operators, respectively,
and  $\gamma_{CE}$ describes the mixing of the color operator 
into the EDM operator. One has 
\begin{equation}
\gamma_E \, = \, \gamma_{CE} \, = \frac{32}{3} \, \; , 
\; \gamma_C  \, = \frac{28}{3} \, ,
\end{equation}
and 
\begin{equation}
\kappa_i \, = \, \frac{\gamma_i}{2\beta_0} \; \, ; \; 
\eta \, = \, \frac{\alpha_s(\mu_{high})}{\alpha_s(\mu_{low})} \; \, ; \; 
\beta_0 \, = \,11 - \frac{2 n_f}{3} \, ,
\end{equation}
where $n_f$ is the number of active quark flavors, which is $n_f=6$ above 
the $t$-quark scale and  $n_f=5$ below.
In the present  case one should  do the running in four steps,
 from the big scale $\mu_\Lambda \sim \Lambda$
down to the top scale $\mu_t \sim m_t$ with $\beta_0 = 7$, from the top scale down
to the $b$-quark scale $m_b$ with $\beta_0 = 23/3$, brom the $b$-quark scale 
down to the charm scale $m_c$ with $\beta_0 = 25/3$, and at last from the
 charm scale down to the hadronic scale $\mu_h \sim$ 1 GeV with $\beta_0 =9$.  

Including QCD corrections, I obtain at the hadronic scale $\mu_h$ = 1 GeV:
\begin{equation}
d_d/e \, = \, C_e (\mu_h) \, = \, K_1 \, \eta^{\frac{16}{21}} \, C_E(\mu_\Lambda)
 \, +  \, \left( K_1  \,(\eta^{\frac{16}{21}} - \eta^{\frac{14}{21}})  \, + \,
K_5 \,  \eta^{\frac{14}{21}} \right) \, C_C(\mu_\Lambda) \; .
\label{QCDdd}
\end{equation}
where $\eta \equiv \alpha_s(\mu_\Lambda)/\alpha_s(\mu_t)$, and where
 $K_1$ and $K_5$ takes care of the QCD corrections below $\mu_t$, and 
 are given in eqs. (\ref{QCDK1}) - (\ref{QCDK5}) in the Appendix.
The one loop result should be a good approximation above top mass scale,
 but not for lower scales, {\it i.e.} not below  say, the $b$-quark scale.

\section{SUMMARY and DISCUSSION}

\begin{center}
\begin{figure}[htbp]
\scalebox{0.55}{\includegraphics{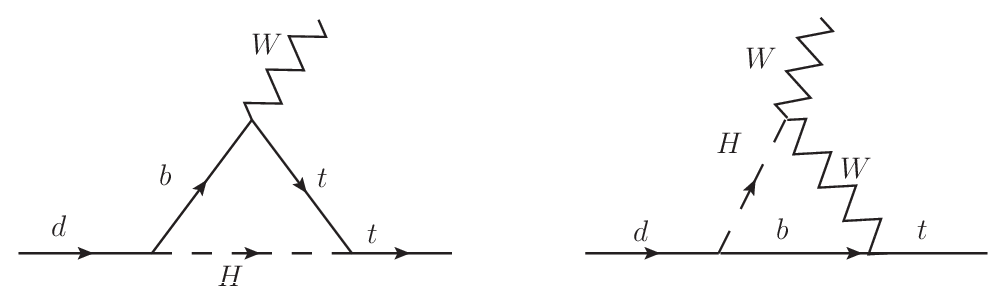}}
\caption{The divergent effective W-loop vertex correction diagram relevant for 
diagrams of section III (left), and the (finite) effective vertex 
correction relevant for diagrams in section IV (right). 
  } 
\label{1loopW}
\end{figure}
 \end{center}

As expected, there are cases where the  considered two loop diagrams 
for the EDMs of $d$- and $u$-quarks diverges. This happens for cases 
in section III where
 the left sub-loop in Fig. \ref{1loopW} is involved, and for diagrams 
where the unphysical Higgs ($\phi_\pm$) is involved  both in sect III and IV.
More specific, the left diagram in Fig. \ref{1loopW} which looks like 
a vertex correction for $d \rightarrow W \, + \, u, c, t$ is 
logarithmically divergent.
 Actually,
this diagram
generates a logarithmic divergent {\it right-handed current} which has no
 match in the SM.
The  diagram at the right in Fig. \ref{1loopW} is convergent, but if
 the $W$-boson  is replaced by an unphysical Higgs $\phi_\pm$,
 when used in two loop diagrams
as in Fig. \ref{2LoopWWHgam}, we obtain logarithmic
divergent  diagrams
due to a momentum dependent vertex, as seen from (\ref{phi}).
 These are numerically relevant 
if the quark $\tilde{q} $ is a top quark.
The dominating  divergent terms in section III and IV are proportional to
$m_t^2$ (-or even $m_t^4$ in one case in section III). It should also 
be noted that
the first and  last diagram in Fig. \ref{2LoopWWHgam} are relevant for 
the EDM of the electron \cite{Altmannshofer:2015qra}. However, in that case 
the divergent terms would be proportional to powers of 
a tiny  neutrino mass, instead of the top-quark mass. 

All contributions (after GIM-cancellation) not proportional
 to $ \xi_t \,  \equiv \, V_{td}^* \, V_{tb}$ are  neglected, using bounds
 on other $Y_R$'s \cite{Blankenburg:2012ex,Harnik:2012pb}, as explaned
 in the preceeding sections. Also all  the contributions for an EDM of
 the $u$-quark can be neglected, for reasons given at the end of the
 sections III and IV.

I have also neglected the $s$-quark contribution
 $d_s$ for the following reason:
The loop functions for the $s$-quark are numerically close to the ones for 
the $d$-quark. The CKM factor 
is bigger, but  $\gamma_s/\gamma_d \simeq 10 ^{-2}$,
 such that the contribution to the result in (\ref{valencedn}) from the
 $d_s$ is of order $5 \%$.
 Thus our final result for the nEDM is simply 
\begin{equation}
d_n \; \simeq \; \rho_d \, d_d \; ,
\label{SimpleEq}
\end{equation}
where  the lattice value of $\rho_d$ is given
in (\ref{valencelatt}).
Using the experimental bound for nEDM in (\ref{dn-bound}),
the result (\ref{eq:AmpTot}) of the present study gives the bound 
\begin{equation}
 | \mbox{Im} \left[Y_R(d \rightarrow b) \cdot
 \frac{V_{td}^* \, V_{tb}}{|V_{td}^* \, V_{tb}|} 
\right] | 
  \le (1.2 \; \,  \mbox{to} \; \, 2.2 ) \times 10^{-3} \; . 
\label{Y1-bound}
\end{equation}
for  values of $\Lambda$  from one  up to seven TeV.

From the mathematical point of view, $\Lambda$ is the quantity
 which regularise the divergent two loop diagrams, while $\Lambda_{NP}$ 
in (\ref{FCNClad}) is introduced as
 a dimensional quantity parametrising the $Y_R$'s and indicates the
 scale of new physics.
But these  scales are expected to be
 of the same order of magnitude.

\begin{center}
\begin{figure}[htbp]
\scalebox{0.55}{\includegraphics{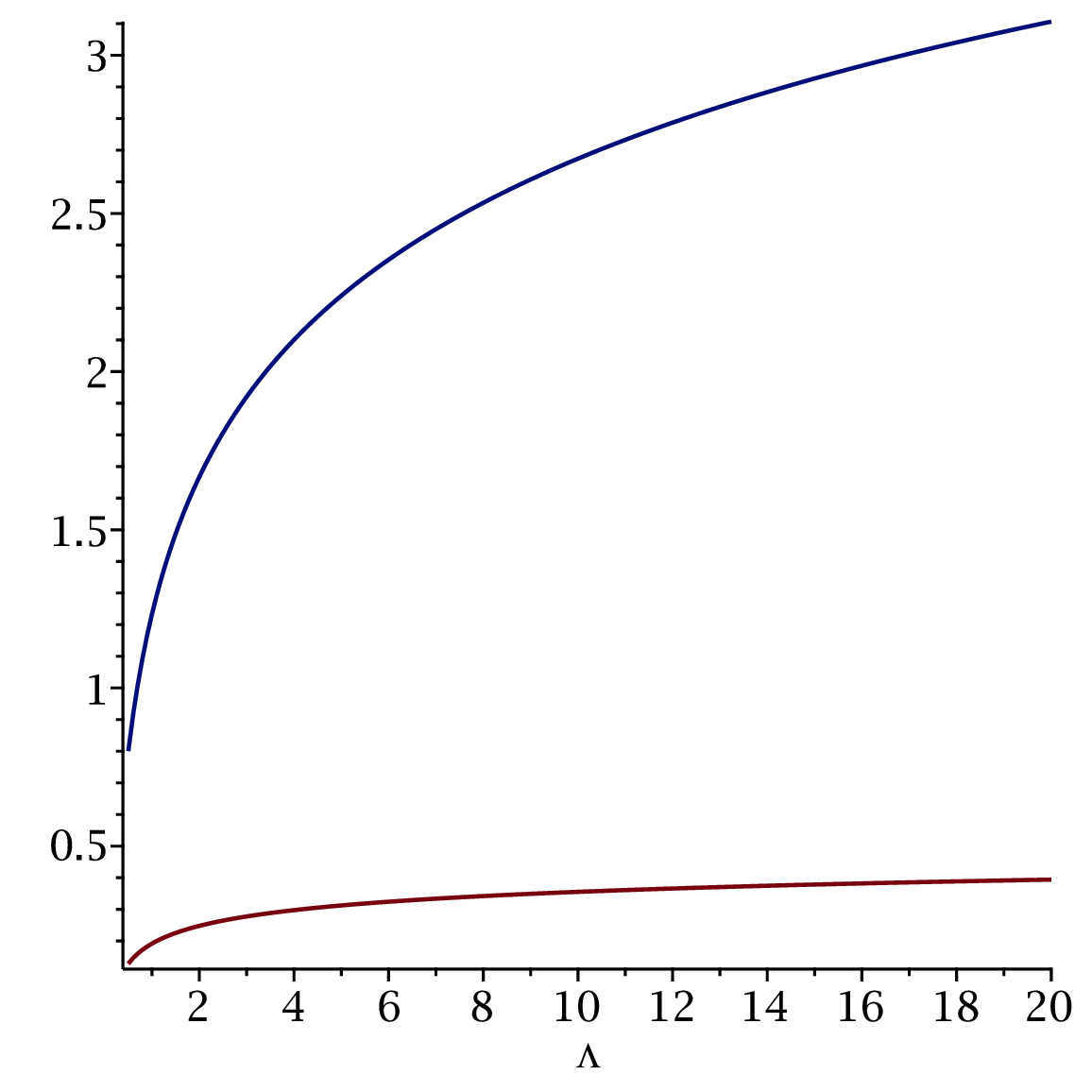}}
\caption{The quantity $N(\Lambda)$, in units $10^{-26}$ cm, 
 as a function  of cut-off $\Lambda$ (in TeV)}
\label{nEDM-numerics}
\end{figure}
 \end{center}

In (\ref{Y1-bound}) I have found a  bound on the imaginary  part of 
the coupling
$Y_R(d \rightarrow b)$ multiplied by the  CKM entry $V_{td}^* V_{tb}$
 (-remenbering that $V_{tb} \simeq 1$). 
Thus the present bound is not directly comparable to the previous bound 
 $1.5 \times 10^{-4}$ on the absolute value of $Y_R(d \rightarrow b)$
given in refs. \cite{Blankenburg:2012ex,Harnik:2012pb}.
But, turning things  around, {\it if} 
 the bound for $Y_R(d \rightarrow b)$ found in 
\cite{Harnik:2012pb} is assumed to be saturated, then
one  can  see how close to the experimental bound on the nEDM
in (\ref{dn-bound}) my value of nEDM might come.
This is illustrated  explicitly as follows:

 Using (\ref{eq:AmpTot}),  the lattice values 
in (\ref{valencelatt}) and absolute value of  $V_{td}^* \, V_{tb}$ from 
 \cite{Olive:2016xmw}, one may write my result for the nEDM in the 
following way
\begin{equation}
d_n/e   \simeq  N(\Lambda) \times 
\left\{ \frac{|Y_R(b \rightarrow d)|}{|Y_R(b \rightarrow d)|_{Bound}} \cdot
\mbox{Im} \left[\frac{Y_R(d \rightarrow b)}{|Y_R(b \rightarrow d)|}
 \cdot \frac{V_{td}^* \, V_{tb}}{|V_{td}^* \, V_{tb}|}\right] \right\}
 \times 10 ^{-26} \,  \mbox{cm} \; ,
\label{dn-f}
\end{equation}
where I have scaled the result with the bound from 
 \cite{Blankenburg:2012ex,Harnik:2012pb}:
 \begin{equation}
| Y_R(d \rightarrow b)| \;  \le \; 1.5 \times 10^{-4} \;  \equiv \; | Y_R(d \rightarrow b)|_{Bound} \, .
\label{YRBound}
\end{equation}
Defining first
\begin{equation}
C_a \, = \, F_2 \, 
Im[Y_R(d \rightarrow b) \, V_{td}^* \, V_{tb}] \, \hat{C}_a \, ,
\end{equation}
for $a=E,C$,
 I further define the function   $N(\Lambda)$,
by the relation
\begin{equation}
 \rho_d \, F_2 \, |Y_R(d \rightarrow b)|_{Bound}  |V_{td}^* \, V_{tb}| 
\,\hat{C_E}(\mu_\Lambda) = \, N(\Lambda)  \, \times 10 ^{-26} \,  \mbox{cm} \, .
\label{NLambda}
\end{equation}
 The function $N(\Lambda)$ is plotted as a
function of $\Lambda$ in Fig. \ref{nEDM-numerics}.

 Now, the {\it maximal value} of 
the parenthesis $\{...\}$  in (\ref{dn-f}) is $= 1$. 
Then, {\it if} the bound for $Y_R(d \rightarrow b)$ is saturated,
 the plot for the function $N_\Lambda$  in Fig. \ref{nEDM-numerics}
shows that when the cut-off $\Lambda$ is stretched up to 20 TeV, the bound for 
nEDM in (\ref{dn-bound}) is reached in the bare case, while the perturbative
 QCD-suppression tells us that the value of the nEDM can at maximum be of 
order one tenth of the experimental bound for $\Lambda$ up to 20 TeV. If the
 bound for $|Y_R(d \rightarrow b)|$ is reduced, and also $\Lambda$ is reduced,
 my value for nEDM will be accordingly smaller.

\section{Conclusion}
 
In conclusion, I have explored the consequenses for the nEDM 
 of having  flavor changing
 Higgs couplings. In the scenary of \cite{Harnik:2012pb,Dorsner:2015mja} 
such couplings might stem from a six dimensional
 non-renormalizable, 
$SU(2)_L \times U(1)_Y$ gauge-invariant  Lagrangian piece proportional to the
 third power of the SM Higgs doublet field, as seen in eq. (\ref{FCNClad}).
While previous analysis  \cite{Blankenburg:2012ex,Harnik:2012pb}
 obtained bound(s) of {\it quadratic} expressions of the 
FCH coupling(s), in the present paper  the analysis is  extended
 to the two loop case for quark EDMs generated by
 a flavor changing Higgs coupling $Y_R(d \rightarrow d)$ to
 {\it first order only}.

I have found and calculated  two loop contributions which  gives a bound for 
the imaginary part of the {\it product} of
$Y_R(d \rightarrow b)$  {\it and} the CKM entry $V_{td}^* V_{tb}$
 (where $V_{tb}$ is known to be very close to one).
This bound cannot be directly compared with the bound 
from  \cite{Blankenburg:2012ex,Harnik:2012pb}, which is on the absolute value.
But even if this bound on the absolute value is saturated, and even  if 
 $\Lambda$ is stretched up to 20 TeV,
 it is seen from Fig. \ref{nEDM-numerics}  that  the value of the
 nEDM can at maximum be of order  one tenth of the
present  experimental bound in (\ref{dn-bound}).

\section{Appendix}

Many loop diagrams are suppressed because of chirality ($P_L \, P_R \, = \, 0$),
 or asymmetric (odd) momentum integration like:
\begin{equation}
\int \dbar \, p  \, f(p^2; \mbox{masses}) \, p^\mu \, =
 \, 0 \; \, .
\label{Antisymint}
\end{equation}

To simplify calculations I use the effective quark propagator in a soft 
electromagnetic field $F$ \cite{Reinders:1984sr}:
\begin{equation}
S_1(k,F) \; = \; (- \frac{e_q}{4}) \;
 \frac{ \left\{(\gamma \cdot k + m_q) \, , \, \sigma \cdot F \right\} }
{(k^2 - m_q^2)^2} \; \; ,
\label{qg-propagator}
\end{equation}
where $k$ is the four momentum and $m_q$ the mass of the quark $q$.
The notation $\{A,B\} \equiv A B + B A$ is used.
Similarly, for emission of a soft photon from a $W$-boson, the
 effective propagator is:
\begin{equation}
(D_1(k,F))^{\alpha \beta}  \; = \; (- \frac{e_W}{4}) \;
 \frac{ 3 \, i \, \left(g^{\mu \alpha}g^{\nu \beta} 
- g^{\mu \beta}g^{\nu \alpha} \right) \, F_{\mu \nu} }{(k^2 - M_W^2)^2} \; \, .
\label{Wg-propagator}
\end{equation}
These effective propagators can be used and are useful when the particles
 in the loop are much bigger that the masses of the external particles.

A typical example for a finite loop integral is
\begin{equation}
T_{\mu \nu} \, = \, \int \frac{\dbar p \, \dbar r \; p_\mu \, r _\nu}
{(p^2 - M_H^2) (p^2 - m_b^2) ((r-p)^2 - M_W^2)(r^2 - M_W^2) (r^2 - m_t^2)^2}.
\, \;   
\label{loopint}
\end{equation}
where $\dbar r \equiv d^4 r/(2 \pi)^4$.
Integrating out momenta, the result of the loop integration gives: 
\begin{equation}
T_{\mu \nu} \, = \, \left(\frac{1}{16\pi^2}\right)^2 \, \frac{g_{\mu \nu}}{4 M_W^2}
\, a(m_t^2, M_W^2, M_H^2) \, ,
\label{loopTens}
\end{equation}
where the dimensionless loop function  $a(m_t^2, \, M_W^2, \, M_H^2)$
 can be written in the compact form
\begin{equation}
a(m_t^2, \, M_W^2, \, M_H^2) \, = \, \int_0^1 d x \, \int_0^{(1-x)} d y \, 
\frac{x y M_W^2}{x(1-x)(M_F^2 - M_H^2)} \left[ \, 1 \, - \,
\frac{M_H^2}{M_F^2 - M_H^2)} \,
 ln(\frac{M_F^2}{M_H^2}) \right]
 \label{loopFunc}
\end{equation}
where the quantity $M_F^2$ depends on masses and Feynman parameters:
\begin{equation}
 M_F^2 \, = \, \frac{M_W^2 \, + \, y (m_t^2 \, -\, M_W^2)}{x(1-x)} \; .
\end{equation}
The expression in (\ref{loopFunc}) can be further found in terms
 of logarithmic and dilogarithmic functions.
Other finite terms are given with formulae as
 (\ref{loopint})-(\ref{loopFunc}), but with masses permuted.
A term with  ultraviolet  divergence appears if $r_\nu$
is replaced by $ p_\nu$  
in the numerator when  doing loop integration in the subloop containing 
the integration over $p$.
This  happens for instance when the $W$-boson is replaced by an 
unphysical Higgs $\phi_\pm$ (the longitudinal $W$-components)
 within Feynman gauge, or if the left diagram in Fig. \ref{1loopW} is involved.
In addition to $a(m_t^2, \, M_W^2, \, M_H^2)$, the loop diagrams in 
section III wil be proportional to $m_t^2/M_W$.

Divergent parts from the first subloop enters as
\begin{equation}
ln(\Lambda^2/R) \; ,
\label{LogR}
\end{equation}
where $R = Q -x(1-x)r^2$, where $Q$ is a quantity depending on masses and 
Feynman paprameters.(for inst $Q= M_W^2 x + M_H^2 y$, where $x$ and $y$
 are Feynman parameters of the first, divergent, subloop). The $ln(R)$ term
 results in a finite term . For example, for third diagram
 (with $W^{\pm} \rightarrow \phi_\pm$, one obtains :
\begin{eqnarray}
S_{3N}^\phi \, = \, - \,   
 \frac{u_t^2}{(u_H-b)} \,\int_0^1 dx \, x(1-x) [ N(1,u_t;B_1) - N(1,u_t;B_0)] 
\, = \, -1.90 \; ,
\label{S3Z}
\end{eqnarray}
where $B_{0,1}$ are given as
\begin{equation}
B_0  \, =
 \; \frac{b + x (u_t - b)}{x (1-x)}
\; ; \; B_1   \, = \,
 \frac{u_H + x (u_t - u_H)}{x (1-x)} \; ,
\label{B01}
\end{equation}
where $u_t$ and $u_H$ are given in (\ref{mass-ratios}), and 
$b \equiv m_b^2/M_W^2$.

\begin{eqnarray}
N(C,A;B) \, \equiv \,  \frac{(A^2-A B)}{C-A}\left( \frac{ln(B)}{A} + 
\frac{ln(\frac{B}{A})}{(B-A)} \right) \nonumber \\   + \, 
\frac{C(C-B)}{(C-A)^2} \{ \frac{1}{2}[ln(B-C)]^2 + dilog(\frac{B}{B-C}) \, - \, 
ln(C) \cdot ln(B-C) \} \, 
\nonumber \\
+ \, \frac{[B C -A(2C-A)]}{(C-A)^2} \{ \frac{1}{2}[ln(B-A)]^2 +
 dilog(\frac{B}{B-A}) \, - \, 
ln(A) \cdot ln(B-A) \} \, .
\label{N-funct}
\end{eqnarray}
Further, there is a non-logarithimic 
finite term (not in $ln(R)$ in eq. (\ref{LogR})):
For other divergent diagrams one has similar expressions with permuted masses.

The dilogarithmic function  is in my case  defined as
\begin{equation}
dilog(z) \, \equiv \, \int_1^z dt \frac{ln(t)}{(1-t)} \; = 
\, \int_0^1 \frac{dx}{x} ln(1-(1-z)x) \, =  
\, Li_2(1-z) \, .
\label{dilog}
\end{equation}

The QCD correction factors in (\ref{QCDdd}) are 
\begin{equation}
K_1 \, = \, (\frac{\alpha_s(\mu_c)}{\alpha_s(\mu_h)})^{\frac{16}{27}}
\, (\frac{\alpha_s(\mu_b)}{\alpha_s(\mu_c)})^{\frac{16}{25}}
\, (\frac{\alpha_s(\mu_t)}{\alpha_s(\mu_b)})^{\frac{16}{23}} \; ,
\label{QCDK1}
\end{equation}
\begin{equation}
K_2 \, = \, 8 \left( (\frac{\alpha_s(\mu_c)}{\alpha_s(\mu_H)})^{\frac{16}{27}}
\, (\frac{\alpha_s(\mu_b)}{\alpha_s(\mu_c)})^{\frac{16}{25}}
\, [(\frac{\alpha_s(\mu_t)}{\alpha_s(\mu_b)})^{\frac{16}{23}} \, - \, 
(\frac{\alpha_s(\mu_t)}{\alpha_s(\mu_b)})^{\frac{14}{23}}] \right)  \; ,
\label{QCDK2}
\end{equation}
\begin{equation}
K_3 \, = \, 8 \left( (\frac{\alpha_s(\mu_c)}{\alpha_s(\mu_h)})^{\frac{16}{27}}
\, [(\frac{\alpha_s(\mu_b)}{\alpha_s(\mu_c)})^{\frac{16}{25}} \, - \,
(\frac{\alpha_s(\mu_b)}{\alpha_s(\mu_c)})^{\frac{14}{25}}]
\, (\frac{\alpha_s(\mu_t)}{\alpha_s(\mu_b)})^{\frac{16}{23}} \right) \; ,
\label{QCDK3}
\end{equation}
\begin{equation}
K_4 \, = \, 8 \left([(\frac{\alpha_s(\mu_c)}{\alpha_s(\mu_h)})^{\frac{16}{27}} \, -
\, (\frac{\alpha_s(\mu_c)}{\alpha_s(\mu_H)})^{\frac{14}{27}} ]
\, (\frac{\alpha_s(\mu_b)}{\alpha_s(\mu_c)})^{\frac{16}{25}}
\, (\frac{\alpha_s(\mu_t)}{\alpha_s(\mu_b)})^{\frac{16}{23}} \right) \; ,
\label{QCDK4}
\end{equation}
and 
\begin{equation}
K_5 \, = \, K_2 \, + \, K_3 \, + \, K_4  \; .
\label{QCDK5}
\end{equation}
One could consequently stick to one-loop values for $\alpha_s(\mu)$ at the 
various scales. However, I have used a hybrid version, taking into acount
 higher loop effects (see for example \cite{Deur:2016tte}) which are important
 below $\mu_b = m_b$, say. Then I have used $\alpha_s(\mu_t) =0.109$, 
$\alpha_s(\mu_b) = 0.23$, $\alpha_s(\mu_c) = 0.40$, 
and $\alpha_s(\mu_h) = 0.52$.

{\acknowledgments}
 I am grateful to Svjetlana Fajfer for suggesting these calculations and for
 valuable discussions. Useful comments by Lluis Oliver and
Ivica Picek are also acknowledged.

I am supported in part by the Norwegian
  research council (via the HEPP project).

\bibliographystyle{unsrt}

\end{document}